\pdfoutput=1

\documentclass[11pt]{article}

\usepackage{emnlp2021}

\usepackage{times}
\usepackage{latexsym}

\usepackage[T1]{fontenc}

\usepackage[utf8]{inputenc}

\usepackage{microtype}

\usepackage{hyperref}
\usepackage{graphicx}
\usepackage{multirow}
\usepackage{appendix}
\usepackage{wrapfig,lipsum,booktabs}
\usepackage{color}
\usepackage{url}
\usepackage{subcaption}

\usepackage{amsmath,amsfonts,bm}

\def\eqref#1{equation~\ref{#1}}

\def\1{\bm{1}}

\DeclareMathAlphabet{\mathsfit}{\encodingdefault}{\sfdefault}{m}{sl}
\SetMathAlphabet{\mathsfit}{bold}{\encodingdefault}{\sfdefault}{bx}{n}

\title{CTAL: Pre-training Cross-modal Transformer for Audio-and-Language Representations}

\author{Hang Li, Yu Kang, Tianqiao Liu, Wenbiao Ding, Zitao Liu \\
        TAL Education Group, Beijing, China \\
        \texttt{\{lihang4,kangyu,liutianqiao,dingwenbiao,liuzitao\}@tal.com} \\}

\begin{document}
\maketitle
\begin{abstract}

Existing audio-language task-specific predictive approaches focus on building complicated late-fusion mechanisms. However, these models are facing challenges of overfitting with limited labels and low model generalization abilities. In this paper, we present a \textbf{C}ross-modal \textbf{T}ransformer for \textbf{A}udio-and-\textbf{L}anguage, i.e., CTAL, which aims to learn the intra-modality and inter-modality connections between audio and language through two proxy tasks on a large amount of audio-and-language pairs: masked language modeling  and masked cross-modal acoustic modeling. After fine-tuning our pre-trained model on multiple downstream audio-and-language tasks, we observe significant improvements across various tasks, such as, emotion classification, sentiment analysis, and speaker verification. On this basis, we further propose a specially-designed fusion mechanism that can be used in fine-tuning phase, which allows our pre-trained model to achieve better performance. Lastly, we demonstrate detailed ablation studies to prove that both our novel cross-modality fusion component and audio-language pre-training methods significantly contribute to the promising results. The code and pre-trained models are available at \url{https://github.com/Ydkwim/CTAL}.
\end{abstract}

\section{Introduction}
\label{sec:intro}
Speech processing requires the understanding of a set of acoustic and language content, including phonemes, tone, words and semantic meanings. Unlike humans, who can benefit from self-learning through real-world experiences, current speech processing methods are like narrow experts relying heavily on large amount of task-specific human annotations, a small change in the learning problem means that they have to start all over again. In recent years, pre-training for single modality of natural language processing (NLP) and of audio signal processing are widely explored due to the ease-of-use and competent generalization to various downstream tasks. 

In the field of NLP, pre-trained models, such as BERT \cite{devlin2018bert}, RoBERTa \cite{liu2019roberta}, XLNet \cite{yang2019xlnet} and GPT2 \cite{radford2019language}, share the same idea of first leveraging large-scale unlabeled corpus to perform contextualized language model (LM) pre-training then fine-tuned to adapt to downstream tasks, such as machine reading comprehension \cite{lai2017race}, question answering \cite{rajpurkar2016squad} and natural language inference \cite{bowman2015large}, which has substantially advanced the state-of-the-art results. Following the success of pre-training in NLP, BERT-like models are also applied to audio processing community \cite{schneider2019wav2vec, baevski2019vq, baevski2020wav2vec}, which learn robust audio representations through an audio-style self-supervised context prediction task. 

Despite these influential single-modal methods, for tasks at the intersection of audio and language, such as speech emotion classification \cite{livingstone2018ryerson,busso2008iemocap}, speaker verification \cite{panayotov2015librispeech} and sentiment analysis \cite{zadeh2018multi}, large-scale pre-training for the modality-pair of audio and language is barely explored. The previous attempt is to train task-specific experts upon the concatenation of language representations and audio representations in a late fusion manner \cite{ramirez2011modeling, glodek2011multiple, zadeh2017tensor, yoon2019speech, yoon2018multimodal, xu2019learning}, without any generic audio-and-language pre-training. These task-specific experts will suffer from overfitting problem when trained with limited data. Also, due to the heterogeneities across language and audio modalities, late fusion of high-level representations lacks surface-level cross-modal alignment and complementation during pre-training phase.

Motivated by these, we propose CTAL, a pre-trainable generic representation for audio-and-language and has shown its strong performance on three established audio-and-language tasks - emotion classification \cite{busso2008iemocap}, sentiment analysis \cite{zadeh2018multi} and speaker verification \cite{panayotov2015librispeech}. We propose multi-modal Transformer as our backbone model, which consists of two modules, a language stream encoding module which adapts word as input element, and a text-referred audio stream encoding module which accepts both frame-level Mel-spectrograms and token-level output embeddings from the language stream encoding module as input elements. In order to learn both intra-modality and inter-modality connections, we pre-train our model with two tasks: (1) masked language modeling. (2) masked cross-modal acoustic modeling. Different from single-modality pre-training (e.g., Masked Acoustic Modeling (MAM) in MOCKINGJAY \cite{liu2020mockingjay}), our cross-modal pre-training enables our model to reconstruct masked audio features from both intra-modality and inter-modality information. On the basis of our pre-trained model, a regularization term based on feature orthogonality is introduced during model fine-tuning stage, which is designed to ensure that features of different modalities provide information from different perspectives, and it should be noted that this orthogonal regularization mechanism is general and not limited to audio-language tasks.

The main contributions of our paper are listed as follows:
\begin{itemize}
    \vspace{-0.2cm}
    \item We present CTAL, a pre-training framework for strong audio-and-language representations with Transformer, which are helpful in learning both intra-modality and inter-modality connections. To the best of our knowledge, we are the first to introduce the pre-training cross audio and language modality.
    \vspace{-0.2cm}
    \item We propose a novel cross-modality fusion mechanism during fine-tuning stage, which forces our pre-trained model learn composite features from different views.
    \vspace{-0.2cm}
    \item Comprehensive empirical evidence demonstrates that our CTAL achieves state-of-the-art results on various downstream speech understanding tasks, such as emotion classification, sentiment analysis, and speaker verification. We conduct detailed ablation studies and analysis to prove the effectiveness of our model components and our pre-training strategies.
\end{itemize}







\section{Related Work}
\label{sec:related}
\subsection{Self-Supervised Uni-modal Pre-training}
There has been a long interest in NLP around self-supervised representation learning. Previous works have explored alternative approaches to improve word embedding \cite{mikolov2013efficient, le2014distributed, pennington2014glove}, which is a low-level linguistic representation. After that, pre-trained NLP models based on multi-layer Transformers, such as BERT \cite{devlin2018bert}, RoBERTa \cite{liu2019roberta}, XLNet \cite{yang2019xlnet} and GPT2 \cite{radford2019language}, benefit from context-sensitive representation learning on large-scale corpus, show significant improvements in various downstream language understanding tasks. Self-supervised learning in audio signal processing has also shown increasing promise. Following BERT, many approaches \cite{jiang2019improving, liu2020tera, liu2020mockingjay, chi2020audio} are proposed to learn high level acoustic representations rather than surface features such as log Mel-spectrograms or waveform, which can reveal the abundant information within audio signals.

\subsection{Multimodal Pre-training}
While pre-training for audio-and-language representations has rarely been studied, several attempts have been made to pre-train models for vision-and-language tasks on Visual Question Answering \cite{antol2015vqa} and Visual Commonsense Reasoning \cite{zellers2019recognition} datasets. In general, these vision-and-language pre-training methods can be divided into two categories, according to their different encoder architectures. (a) Prior works like ViLBERT \cite{lu2019vilbert} and LXMERT \cite{tan2019lxmert}, apply two single-modal networks to encode input text and images respectively and adapt cross-modal interactions in a symmetric fusion manner. (b) The other category of pre-training frameworks like VisualBert \cite{li2019visualbert} and Unicoder-VL \cite{li2020unicoder}, concatenate vision and language features as a unified single-stream input and utilize a universal encoder to learn joint multimodal representations. 

To be noticed, transfer above algorithms directly from vision-and-language to audio-and-language field is facing challenges, including: (1) unified architecture is not suitable for audio-language modalities, since both text and audio-waveform are generally long sequences, and cross-modal aggregation at the very beginning phase with Transformer self-attention mechanism will lead to higher computational complexity; (2) speech audio is more informative than language text, which contains both semantic information of speech text and personal feelings of speakers. Thus, it is not suitable to apply the symmetric cross-modal fusion modules proposed in prior audio-and-language pre-training researches. Based on these facts, we design our backbone model with a language stream encoding module and a text-referred audio stream encoding module, which allow necessary intra- and inter-modality connections during pre-training with less computational cost.



The closest work to our approach is from \citet{haque2019audio} and our approach differs from it in two ways: (1) we use a more explicit, multi-component design for the cross-modality connections (i.e., with a text-referred audio stream encoding module and a novel cross-modality fusion component); (2) we employ different pre-training tasks which accept both text and audio frames as input to conduct contextualized masked language modeling and masked cross-modal acoustic modeling tasks. While previous work only adapts audio as input and formulates a multitask learning problem by reconstructing linguistic and acoustic features from a hidden speech embedding during pre-training.

\section{Approach}
\label{sec:approach}
In this section, we first present our cross-modal pre-training framework CTAL, including details of text and audio preprocessing and encoding module for separate modalities. Then we present our pre-training tasks. In the end, we propose our novel fusion mechanism which can be utilized in the fine-tuning stage. Following conventions, we apply bold upper case letters to represent matrices and bold lower case letters to represent vectors. 


\subsection{CTAL}

\begin{figure*}[!hbpt]
    \begin{center}
    \includegraphics[width=0.93\textwidth]{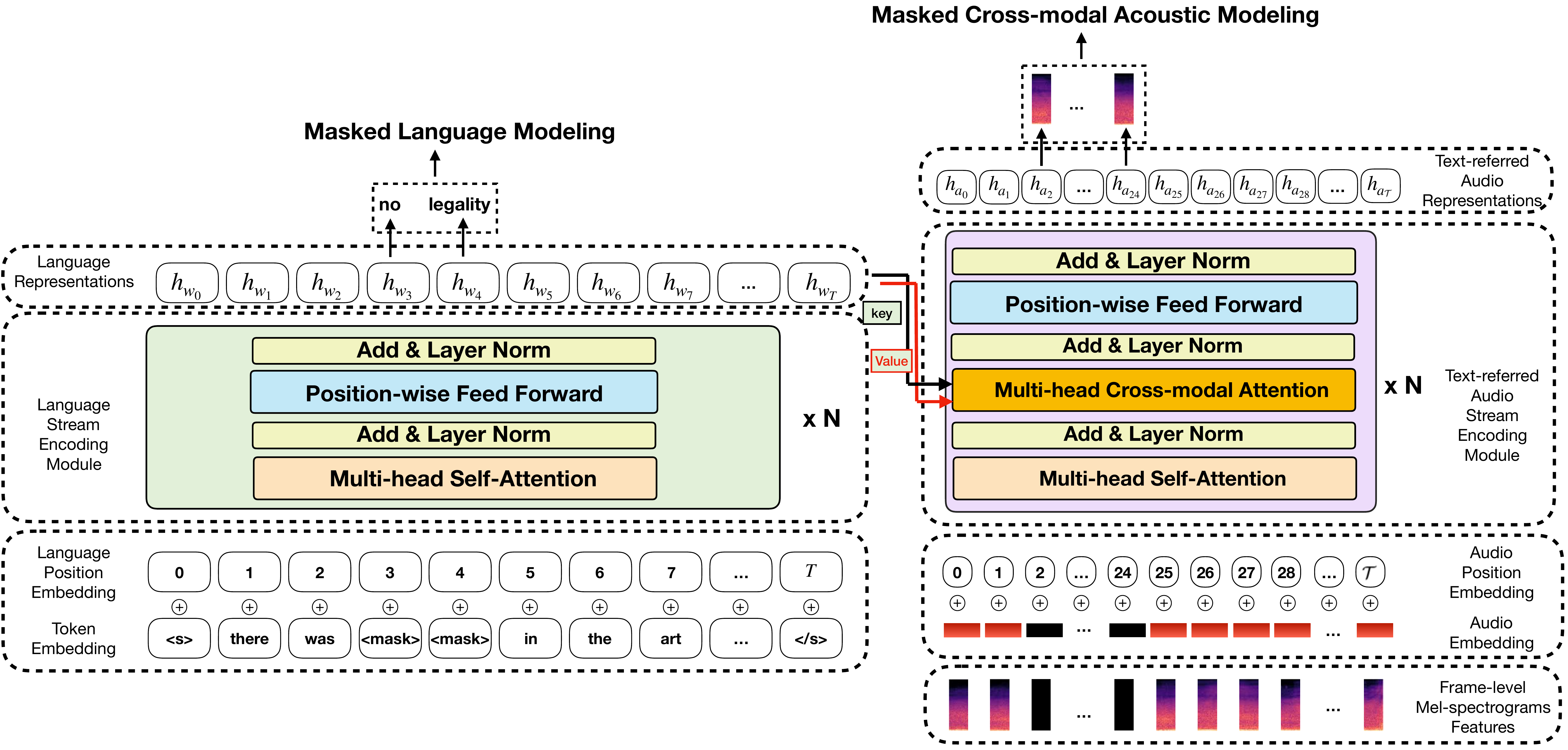}
    \end{center}
    \vspace{-0.25cm}
    \caption{The proposed CTAL pre-training framework.}
    \label{fig:model}
    \vspace{-0.25cm}
\end{figure*}

\label{sec:ctal}
We build our cross-modal Transformer by extending the original Transformer \cite{vaswani2017attention} into the multimodal paradigm. As shown in Figure~\ref{fig:model}, CTAL takes audio sequence and its corresponding text as the input. Each audio is represented as a sequence of frames, and each text is represented as a sequence of tokens. Then we encode the input to the linguistic embedding and audio embedding, and feed them into a text encoding module and a text-referred audio encoding module respectively to generate final language representations and text referred audio representations. Following the formula and notations proposed by \citet{vaswani2017attention}, we adapt $\mathbf{Q}$, $\mathbf{K}$ and $\mathbf{V}$ as queries, keys and values for attention mechanism, MultiHead($\mathbf{Q}$, $\mathbf{K}$, $\mathbf{V}$) as multi-head attention, FFN($\mathbf{X}$) as position-wise feed forward networks and LayerNorm($\mathbf{X}$) as layer normalization. Next, we describe the components in detail.

\subsubsection{Input Embeddings}
\label{sec:input}
\textbf{Linguistic Embedding} \indent To encode any input text with a modest size (30K units) of subword vocabulary, we follow the text preprocessing of RoBERTa, which tokenizes each input text $w=\{w_0, w_1, ..., w_T\}$ with byte-level Byte-Pair Encoding (BBPE) \cite{radford2019language}. Besides, we also add the special tokens <s> and </s> to represent start and end token as the common practice, and $T$ is the total length of input tokens. Then we sum up each token embedding and its corresponding position embedding as the final input token embeddings $\{\mathbf{e}_{w_0}, \mathbf{e}_{w_1}, ..., \mathbf{e}_{w_T}\}$ for language modality.

\noindent \textbf{Audio Embedding} \indent The input audio signal is first transformed into frames of width 50ms and step 12.5ms. Then the 80 dimension Mel-spectrograms are extracted from each frame and concatenated with their first order derivatives, making the feature dimension to 160. In this way, the raw signal is converted into sequence of frame-level acoustic surface features $\{a_0, a_1, ..., a_\mathcal{T}\}$, where $\mathcal{T}$ is the total number of frames. For simplicity, we denote this audio feature sequence as input acoustic features after this section. At final, we feed these surface features to a projection layer and add them with the position embeddings to obtain the input audio embeddings $\{\mathbf{e}_{a_0}, \mathbf{e}_{a_1}, ..., \mathbf{e}_{a_\mathcal{T}}\}$ for audio modality.

\subsubsection{Text Encoding Module}
\label{subsec:text-encoder}
As shown in Figure~\ref{fig:model}, we apply the original Transformer encoder to language stream inputs, each language stream encoding layer consists of one multi-head self-attention sublayer and one position-wise feed forward sublayer. We stack $N$ such language encoding layer in our text encoding module implementation, using the output of $(k-1)$-th layer as the input to the $k$-th layer, and we initialize $\mathbf{H}_w^0$ with $\{\mathbf{e}_{w_0}, \mathbf{e}_{w_1}, ..., \mathbf{e}_{w_T}\}$. We obtain our language representations for $k$-th layer with the followings:
\begingroup\makeatletter\def\f@size{10}\check@mathfonts
\def\maketag@@@#1{\hbox{\m@th\large\normalfont#1}}%
\begin{align*}
\mathbf{\hat{H}}_w^k &= MultiHead(\mathbf{H}_w^{k-1}, \mathbf{H}_w^{k-1}, \mathbf{H}_w^{k-1})
\\
\mathbf{\tilde{H}}_w^k &= LayerNorm(\mathbf{\hat{H}}_w^k + \mathbf{H}_w^{k-1})
\\
\mathbf{H}_w^k &= LayerNorm(FFN(\mathbf{\tilde{H}}_w^k)+\mathbf{\tilde{H}}_w^k)
\end{align*}
\endgroup
We get the final output $\mathbf{H}_w^{N} \in \mathbb{R}^{T \times d_w}$ from our language stream encoding module, where $d_w$ denotes the hidden size of the language stream representations. The first token of every text sequence is always a special start token (<s>), and the final hidden state corresponding to this token is always used as the aggregated text sequence representation for classification tasks.

\subsubsection{Text-Referred Audio Encoding Module}
\label{subsec:audio-encoder}
For text-referred audio encoding module, we first initialize hidden representations $\mathbf{H}_a^0$ with $\{\mathbf{e}_{a_0}, \mathbf{e}_{a_1}, ..., \mathbf{e}_{a_\mathcal{T}}\}$, and pass them to a stack of $N$ text-referred audio encoding layers to acquire the final audio stream representations $\mathbf{H}_a^N$.

Our text-referred audio encoding module is different from the original Transformer decoder by modifying two kinds of multi-head attention mechanism. Firstly, in order to learn the bi-directional intra-modality representation for audio, we get rid of the future mask in the masked multi-head self-attention. Specifically for $l$-th layer: 
\begingroup\makeatletter\def\f@size{10}\check@mathfonts
\def\maketag@@@#1{\hbox{\m@th\large\normalfont#1}}%
\begin{align*}
\mathbf{\hat{H}}_a^l &= MultiHead(\mathbf{H}_a^{l-1}, \mathbf{H}_a^{l-1}, \mathbf{H}_a^{l-1})
\\
\mathbf{\tilde{H}}_a^l &= LayerNorm(\mathbf{\hat{H}}_a^l + \mathbf{H}_a^{l-1})
\end{align*}
\endgroup
Secondly, we apply multi-head cross-modal attention which accepts the final language stream representations as keys and values in each layer to apply the inter-modality interactions:
\begingroup\makeatletter\def\f@size{10}\check@mathfonts
\def\maketag@@@#1{\hbox{\m@th\large\normalfont#1}}%
\begin{align*}
\mathbf{\bar{H}}_a^l &= MultiHead(\mathbf{\tilde{H}}_a^l, \mathbf{H}_w^N, \mathbf{H}_w^N)
\\
\mathbf{\ddot{H}}_a^l &= LayerNorm(\mathbf{\bar{H}}_a^l+\mathbf{\tilde{H}}_a^l)
\\
\mathbf{H}_a^l &= LayerNorm(FFN(\mathbf{\ddot{H}}_a^l)+\mathbf{\ddot{H}}_a^l)
\end{align*}
\endgroup
Finally, we obtain the text-referred audio representation of $N$-th layer $\boldsymbol{H}_{a}^{N} \in \mathbb{R}^{\mathcal{T} \times d_{a}}$, where $d_{a}$ denotes the hidden size of the audio stream representations.

\subsection{Pre-training Tasks}
\label{sec:pretraining_task}
\subsubsection{Masked Language Modeling (MLM)}
\label{sec:mlm}
For language stream, we take the masked language modeling task for language intra-modality learning. As shown in Figure~\ref{fig:model}, in MLM, the task setup is almost the same as RoBERTa \cite{liu2019roberta}, we dynamically mask out the input tokens with a probability of 15\%. Masked tokens are replaced with a special <mask> token 80\% of the time, a random token 10\%, and unaltered 10\%. The goal of MLM is to predict these masked tokens based on the observed tokens. To be noticed, we do not introduce acoustic information for masked token prediction, since semantic information of language text can be well captured through language input only. And introducing cross-modal inputs is redundant,  which we demonstrate through our later ablation study.



\subsubsection{Masked Cross-modal Acoustic Modeling (MCAM)}
For audio stream, we propose MCAM to train the text-referred audio representations. Prior research by \citet{baevski2020wav2vec} indicates that the performance of acoustic pre-trained models on downstream tasks is improved with the increment in size of continuous masked frames during pre-training phase. However, due to the complexity of audio signals, the long-term dependencies in audio sequences is hard to be captured with acoustic features alone. To mitigate that problem, we propose MCAM to capture effective information of audio through learning both intra- and inter-modality connections between audio and language.


To implementation MCAM, we first split the audio in separate segments according to $C_{num}$ consecutive frames per segment, where $C_{num}$ is uniformly sampled from $20$ to $50$. Then we randomly select 15\% of these segments and for each of them, we mask it all to zero 80\% of the time, replace it with the other $C_{num}$ randomly selected frames within the audio 10\% of the time, and keep it unchanged for the remaining cases. In this manner, we prevent the model exploiting local smoothness of acoustic frames and the model is required to make inference based on global information rather than local messages. Finally, the goal is to reconstruct these masked acoustic features $\{a_i|i\in\mathcal{T}_{mask}\}$ based on the remaining acoustic features and the language stream prompt, by minimizing the L1 loss between the original masked acoustic features and the predicted ones.


Overall, our final pre-training objective is to minimize the sum of the losses above.

\subsection{Fine-Tuning CTAL}
\label{sec:finetune}
CTAL is designed to be a generic pre-training model for various audio-language tasks. It is relatively simple to fine-tune CTAL for various downstream tasks with just one additional output layer. To further combine information from different modalities, we propose a novel and flexible fusion mechanism at fine-tuning stage. We denote $\boldsymbol{H}_{w}^N \in \mathbb{R}^{T\times d}$ and $\boldsymbol{H}_{a}^N \in \mathbb{R}^{\mathcal{T} \times d}$ as the final representation from text encoding module and text-referred audio encoding module, and we assume that both modules have the same hidden size $d$. 

To fine-tune on speech understanding tasks, we are required to represent the input sequence (for both language and audio stream) to a compressed hidden vector. Following the idea from \citet{wang2018polyphonic}, which proves that max pooling mechanism has a tendency to make too many false negatives while attention pooling mechanism prefers making too many false positives, we come up with both Attention-Pooling layer and Max-Pooling layer to let them complement each other. After applying Attention-Pooling and Max-Pooling to audio stream final representations $\boldsymbol{H}_{a}^N$, We obtain $\mathbf{h}^{attn}_a \in \mathbb{R}^d$ and $\mathbf{h}^{max}_a \in \mathbb{R}^d$ respectively.\begingroup\makeatletter\def\f@size{10}\check@mathfonts
\def\maketag@@@#1{\hbox{\m@th\large\normalfont#1}}%
\begin{align*}
\mathbf{h}^{attn}_a &=\mathit{Softmax}(\mathbf{v}^{attn}_a \cdot \mathrm{tanh}(\mathbf{W}^{attn}_a \cdot \boldsymbol{H}_{a}^N)) \cdot \boldsymbol{H}_{a}^N \nonumber
\\
\mathbf{h}^{max}_a &= \mathit{MaxPool}(\boldsymbol{H}_{a}^N) \nonumber
\end{align*}
\endgroup

\noindent where $\mathbf{v}^{attn}_a$ and $\mathbf{W}^{attn}_a$ are trainable parameters for audio side Attention-Pooling layer.

As discussed in Section~\ref{subsec:text-encoder}, for language stream, we adapt the final hidden state of the start token $\mathbf{h}_{w0} \in \mathbb{R}^d$ as the aggregated text sequence representation $\mathbf{h}^{attn}_w$ for Attention-Pooling, and we conduct additional Max-Pooling for text stream output $\boldsymbol{H}_{w}^N$ to obtain $\mathbf{h}^{max}_w$.

Then we fuse the aggregated sequence representations from two modalities as follows:
\begingroup\makeatletter\def\f@size{10}\check@mathfonts
\def\maketag@@@#1{\hbox{\m@th\large\normalfont#1}}%
\begin{equation}
\mathbf{h}^{fuse} = (\mathbf{h}^{attn}_{a}+\mathbf{h}^{attn}_{w}) \ \oplus \ (\mathbf{h}^{max}_{a}+\mathbf{h}^{max}_{w}) \nonumber
\end{equation}
\endgroup
where $\oplus$ denotes the vector concatenation, and the final hidden state $\mathbf{h}^{fuse}$ is always used as the audio-and-language representation for classification tasks.


\subsubsection{Orthogonal Regularization}
One key characteristic of multimodal learning is the generated representations of different modality are supposed to depict a sample from different point of views. In order to encourage the two modules to get representations from different perspectives rather than similar characteristic. In addition to the loss function inherent to each task, we also introduce a regularization term which is minimized simultaneously with the objective to achieve the representations orthogonality during fine-tuning stage:
\begingroup\makeatletter\def\f@size{10}\check@mathfonts
\def\maketag@@@#1{\hbox{\m@th\large\normalfont#1}}%
\begin{equation}
    \mathcal{L}_{\mathrm{Orth}} = \frac{|{\mathbf{h}^{attn}_{a}}^{T} \mathbf{h}^{attn}_{w}|}{\left \| \mathbf{h}^{attn}_{a} \right \|  \left \| \mathbf{h}^{attn}_{w} \right \| } + \frac{|{\mathbf{h}^{max}_{a}}^{T} \mathbf{h}^{max}_{w}|}{\left \| \mathbf{h}^{max}_{a} \right \|  \left \| \mathbf{h}^{max}_{w} \right \| } \nonumber
\end{equation}
\endgroup

\section{Experimental Setup and Result}
\label{sec:experiment}
In this section, we present CTAL pre-training details and fine-tuning results on three downstream audio-and-language tasks.

\subsection{Pre-training Details}
\label{sec:pretrain_details}
We pre-train our CTAL on the public dataset LibriSpeech \cite{panayotov2015librispeech}, which is a dataset of reading English speech, including both audio recordings and corresponding authorized transcripts. It has 7 subsets in total (train-clean-100, train-clean-360, train-other-500, dev-clean, dev-other, test-clean, test-other). The subset with "clean" in its name contains audios with higher recording quality, while the other subsets have relatively lower quality recordings. We use all three training subsets for pre-training, including approximately 960 hours of speech and 280k utterances. 

Following \citet{radford2019language}, we consider training a BBPE tokenizer on the LibriSpeech corpus with additional special tokens (<s>, </s>, <mask>, <pad>) as our language stream tokenizer, and we tokenize the input text into token sequence as described in Section~\ref{sec:input}. For audio stream, we use Librosa \cite{mcfee2015librosa}, which is a well-established audio analysis Python package, to extract the 160-dimension input acoustic feature for each frame as described in Section~\ref{sec:input}. For the pre-train model architecture, we denote the number of layers (i.e., language stream encoding layer and text-referred audio stream encoding layer) as N, the number of self-attention heads as A, and the number of hidden size as H. we primarily report results on two model sizes:$\mathbf{CTAL}_{\mathbf{BASE}}$ (N=3, A=12, H=768) and $\mathbf{CTAL}_{\mathbf{LARGE}}$ (N=6, A=12, H=768). The total number of parameters for $\mathbf{CTAL}_{\mathbf{BASE}}$ is 60M and 110M for $\mathbf{CTAL}_{\mathbf{LARGE}}$. More implementation details in Appendix \ref{sec:pre-train details}

\subsection{Fine-tuning on Downstream Tasks}
\label{sec:fine-tune downstream}
We transfer our pre-trained CTAL model to a set of three established speech understanding tasks, with simple and necessary modifications on the output layers, loss function and training strategy.

\subsubsection{Emotion Classification}
In emotion classification task, given a speech clip, the model is asked to predict which emotion category the speech is belonging to. Here, we conduct experiments on the widely-used dataset IEMOCAP \cite{busso2008iemocap}. The dataset was recorded from ten actors, divided into five sessions, and each session has dialogues between two speakers with different genders. The dataset contains audio, transcriptions, and video recordings, we only use audio and transcriptions in our study. The recorded dialogues have been sliced into utterances and labeled in 10 categories by three annotators and utterances without any text content are filtered out in our experiment. For consistent comparison with previous works, we follow the settings with \citet{xu2019learning}, which use four emotions (angry, happy, neutral and sad) for classification and perform 5-fold cross-validation over sessions, where each session is used as the test set in turn and remaining as training dataset. We adopt two widely used metrics for evaluation: weighted accuracy (WA) that is the overall classification accuracy and unweighted accuracy (UA) that is the average recall over all four classes. We report the averaged WA and UA over the 5-fold cross-validation experiments, and higher WA and UA results represent better model performances.

\begin{table}
\footnotesize
\begin{center}
\begin{tabular}{l|c|c}
\toprule
Methods    & WA  $\uparrow$   & UA  $\uparrow$   \\
\midrule
LSTM\_alignment \cite{xu2019learning} & 0.6900 & 0.7014 \\ 
MRDE \cite{yoon2018multimodal}           & 0.6702 & 0.6764 \\ 
MHA \cite{yoon2019speech}            & 0.6780 & 0.6880 \\
\midrule
$\mathrm{CTAL_{BASE}}$      & 0.7286 & 0.7370 \\
$\mathrm{CTAL_{LARGE}}$     & \bf{0.7395} & \bf{0.7463} \\
\bottomrule
\end{tabular}
\end{center}
\vspace{-0.2cm}
\caption{Comparison to the SOTA methods on the IEMOCAP dataset.}
\label{tab:exp_result:emotion classification}
\vspace{-0.1cm}
\end{table}

To fine-tune on IEMOCAP, we represent the input sequence (for a pair of audio and text) as described in Section~\ref{sec:pretrain_details}, and use the final hidden vector $\mathbf{h}^{fuse}$ as the audio-and-language representation. The only new parameters introduced during fine-tuning are classification layer weights $\mathbf{W} \in \mathbb{R}^{4 \times d}$ and CTAL fine-tuning is driven by the cross-entropy loss between the predicted class and the gold label. More implementation details in Appendix \ref{sec:fine-tune details} We select multiple models that claim to achieve SOTA results on IEMOCAP dataset as our baselines, and to be noticed, previous methods are specifically designed for the task with no pre-training stage. See details in Appendix \ref{sec:fine-tune baseline}

Table~\ref{tab:exp_result:emotion classification} presents our experimental results on IEMOCAP dataset. Since some prior works experiment with different train/test split, we reimplement baseline models with their published codes\footnote{MDRE:\url{https://github.com/david-yoon/multimodal-speech-emotion.git}}\footnote{LSTM\_alignment:\url{https://github.com/didi/delta}}. Both $\mathrm{CTAL_{BASE}}$ and $\mathrm{CTAL_{LARGE}}$ outperform all three baselines by a substantial margin, obtaining 3.86\% and 4.95\% respective absolute WA improvement, and 3.56\% and 4.49\% respective absolute UA improvement over the prior state of the art.


\subsubsection{Sentiment Analysis}
The goal of the sentiment analysis task is to predict the degree of positive and negative sentiment. Compared to the emotion classification task, sentiment analysis is a regression task rather than a classification task. We adopt CMU-MOSEI \cite{zadeh2018multi} dataset for evaluation, which contains 23,454 movie review video clips taken from YouTube. We use only audio and corresponding transcriptions as input in our experiments. Each sample in the dataset is labeled with a sentiment score from -3 (strongly negative) to 3 (strongly positive) by human annotators. We follow the same experimental protocol as MuIT \cite{tsai2019multimodal}, with the same train/test data split and the same evaluation metrics, which includes two classification metrics: binary accuracy (i.e., $\rm{Acc_2}$: accuracy over positive/negative sentiments classification), and F1 score, two regression metrics: mean absolute error (MAE), and the Pearson correlation coefficient (Corr) between model's predictions and human annotations. Since the prior top results reported on the CMU-MOSEI dataset are all achieved using all three modalities, so does MulT\footnote{MulT:\url{https://github.com/yaohungt/Multimodal-Transformer}}, we prune the vision-related components in MulT and re-train the model using only audio and text information.

\begin{table}[]
\footnotesize
\begin{tabular}{l|c|c|c|c}
\toprule
Methods & $\rm{Acc_2}$ $\uparrow$ & F1  $\uparrow$   & MAE $\downarrow$   & Corr $\uparrow$  \\ \midrule
MulT       & 0.7966 & 0.8008 & 0.6367 & 0.6292 \\ \midrule
$\mathrm{CTAL_{BASE}}$  & 0.8036 & 0.8055 & 0.6061 & \bf{0.6828} \\
$\mathrm{CTAL_{LARGE}}$ & \bf{0.8077} & \bf{0.8101} & \bf{0.6027} & 0.6809 \\ \bottomrule
\end{tabular}
\vspace{-0.2cm}
\caption{Comparison to the SOTA methods on the CMU-MOSEI dataset.}
\label{tab:exp_result_sentiment}
\vspace{-0.2cm}
\end{table}

During fine-tuning on sentiment analysis, we introduce additional parameters $\mathbf{W} \in \mathbb{R}^{1 \times d}$ to project the final hidden representation $\mathbf{h}^{fuse}$ to the sentiment score, and the model is trained to minimize the L1 Loss between the predicted scores and the gold annotations. The other fine-tune settings over CMU-MOSEI are almost the same as IEMOCAP. As show in Table~\ref{tab:exp_result_sentiment}, we observe improvements across all 4 metrics for CTAL over MulT baseline under both base and large settings.

\subsubsection{Speaker Verification}

Speaker verification focuses on verifying the speaker identity of an utterance through comparing it with the pre-recoded voice print information. In this experiment, we adopt LibriSpeech \cite{panayotov2015librispeech} for evaluation, which includes 292k utterances collected from more than 2,438 speakers. Following the same experiment setting with prior works \cite{wan2018generalized,jung2019rawnet}, we fine-tune our pre-trained model with all training splits (train-clean-100, train-clean-360 and train-other-500), and evaluate it with test-clean part, which contains 40 brand new speakers to the training part. Please note here, although the train set for our speaker verification task is identical with the one we used for pre-training, the speaker identity information and test-clean data are not released during the pre-training process. Thus, it is fair to perform comparisons between our models with other prior works. We add a classifier over the head of fused embeddings $\mathbf{h}^{fuse}$ and adopt cross-entropy loss to fine-tune it. The output size of the classifier is same to the number of unique speakers in train set.

For evaluation, we utilize the representation before classifier as the input audio's identity embedding. And the cosine distance of each paired audio embeddings are used as the indicator for the final decision. Similar to prior studies, we report the Equal Error Rate (EER) as the evaluation metric, and lower EER represents better model performance. We choose two SOTA models as our baselines \cite{wan2018generalized, jung2019rawnet}. See more details in Appendix \ref{sec:fine-tune baseline speaker verification}. The comparison results are shown in Table.~\ref{tab:exp_result_speaker}. From the table, we observe that our $\mathrm{CTAL_{BASE}}$ outperforms GE2E and RawNet by 1.85\% and 0.59\% respectively, and $\mathrm{CTAL_{LARGE}}$ outperforms two baselines by 2.24\% and 0.98\% respectively.

\begin{table}[!hbpt]
\footnotesize
\centering
\begin{tabular}{l|c}
\toprule
Methods    & EER $\downarrow$   \\
\midrule
GE2E \cite{wan2018generalized}       & 0.0379 \\
RawNet \cite{jung2019rawnet}     & 0.0253 \\
\midrule
$\mathrm{CTAL_{BASE}}$ & 0.0194 \\
$\mathrm{CTAL_{LARGE}}$ & \bf{0.0155} \\
\bottomrule
\end{tabular}
\vspace{-0.2cm}
\caption{Comparison to the SOTA methods on the LibriSpeech dataset.}
\label{tab:exp_result_speaker}
\end{table}

\section{Analysis}
\label{sec:analysis}

\subsection{Ablation Studies}

\begin{table*}[!hbpt]
    \tiny
    \centering
    \setlength{\tabcolsep}{1.4mm}{
        \begin{tabular}{@{}ccccccccc|cc|cccc|c@{}}
        \toprule
        \multirow{2}{*}{Settings} & \multirow{2}{*}{MLM} & \multirow{2}{*}{MCAM} & \multirow{2}{*}{\begin{tabular}[c]{@{}c@{}}Orthognal \\ Fusion\end{tabular}} & \multirow{2}{*}{\begin{tabular}[c]{@{}c@{}}Cross-\\ modal \\ Pre-train\end{tabular}} & \multirow{2}{*}{\begin{tabular}[c]{@{}c@{}}Text \\ Outputs\end{tabular}} & \multirow{2}{*}{\begin{tabular}[c]{@{}c@{}}Audio \\ Outputs\end{tabular}} & \multirow{2}{*}{\begin{tabular}[c]{@{}c@{}}Pre-train\\ 960 \\ Hours\end{tabular}} & \multirow{2}{*}{\begin{tabular}[c]{@{}c@{}}Pre-train\\ 360 \\ Hours\end{tabular}} & \multicolumn{2}{c|}{\begin{tabular}[c]{@{}c@{}}Emotion \\ Classification\\ (IEMOCAP)\end{tabular}} & \multicolumn{4}{c|}{\begin{tabular}[c]{@{}c@{}}Sentiment \\ Analysis\\ (MOSEI)\end{tabular}} & \begin{tabular}[c]{@{}c@{}}Speaker \\ Verification\\ (LibriSpeech)\end{tabular} \\ \cmidrule(l){10-16} 
            &  &  &  &  &  &  &  &  & WA $\uparrow$ & UA $\uparrow$ & $\rm{Acc_2}$ $\uparrow$ & F1 $\uparrow$ & MAE $\downarrow$ & Corr $\uparrow$ & EER $\downarrow$ \\ \midrule
        w/o pre-training &  &  & $\surd$ &  & $\surd$ & $\surd$ &  &  & 0.7004 & 0.7110 & 0.7804 & 0.7809 & 0.6654 & 0.6086 & 0.0366 \\ \midrule
        (a) & $\surd$ & $\surd$ & $\surd$ & $\surd$ & $\surd$ & $\surd$ &  & $\surd$ & 0.7262 & 0.7386 & \textbf{0.8127} & \textbf{0.8150} & 0.6050 & 0.6804 & 0.0204 \\ 
        (b) &  & $\surd$ & $\surd$ & $\surd$ & $\surd$ & $\surd$ & $\surd$ &  & 0.7077 & 0.7185 & 0.7834 & 0.7842 & 0.6629 & 0.6096 & 0.0244 \\ 
        (c) & $\surd$ &  & $\surd$ &  & $\surd$ & $\surd$ & $\surd$ & & 0.7080 & 0.7171 & 0.7812 & 0.7809 & 0.6442 & 0.6440 & 0.0327 \\ 
        (d) & $\surd$ & $\surd$ &  & $\surd$ & $\surd$ & $\surd$ & $\surd$ &  & 0.7338 & 0.7444 & 0.7948 & 0.7939 & 0.6035 & \textbf{0.6832} & 0.0176 \\ 
        (e) & $\surd$ & $\surd$ &  & $\surd$ & $\surd$ &  & $\surd$ &  & 0.6497 & 0.6586 & 0.7804 & 0.7795 & 0.6235 & 0.6639 & - \\ 
        (f) & $\surd$ & $\surd$ &  & $\surd$ &  & $\surd$ & $\surd$ &  & 0.7315 & 0.7412 & 0.7989 & 0.7915 & 0.6065 & 0.6750 & 0.0190 \\ 
        (g) & $\surd$ & (MAM) & $\surd$ &  & $\surd$ & $\surd$ & $\surd$ &  & 0.7116 & 0.7270 & 0.7820 & 0.7834 & 0.6323 & 0.6527 & 0.0306 \\ \midrule
        $\mathrm{Mockingjay}$ & & (MAM) & & & & $\surd$ & $\surd$ & & 0.5505 & 0.5672 & 0.6887 & 0.7199 & 0.8056 & 0.3556 & 0.0551 \\
        $\mathrm{RoBERTa}$ & $\surd$ & & & & $\surd$ & & $\surd$ & & 0.6377 & 0.6411 & 0.7451 & 0.7412 & 0.6598 & 0.5760 & - \\ \midrule
        $\mathrm{LXMERT}$ & \multicolumn{2}{c}{(LXMERT)} & $\surd$ & $\surd$ & $\surd$ & $\surd$ & $\surd$ & & 0.7145 & 0.7222 & 0.7749 & 0.7740 & 0.6405 & 0.6430 & 0.0320 \\ 
        $\mathrm{CTAL_{BASE}}$ & $\surd$ & $\surd$ & $\surd$ & $\surd$ & $\surd$ & $\surd$ & $\surd$ &  & 0.7286 & 0.7370 & 0.8036 & 0.8055 & 0.6061 & 0.6828 & 0.0194 \\ 
        $\mathrm{CTAL_{LARGE}}$ & $\surd$ & $\surd$ & $\surd$ & $\surd$ & $\surd$ & $\surd$ & $\surd$ &  & \textbf{0.7395} & \textbf{0.7463} & 0.8077 & 0.8101 & \textbf{0.6027} & 0.6809 & \textbf{0.0155} \\ \bottomrule
        \end{tabular}
    }
    \caption{The results for performing ablation study with $\mathrm{CTAL_{LARGE}}$. Notation (MAM) represents the acoustic stream encoding module is pre-trained with MAM task. The EER is not reported for setting (d) and $\mathrm{RoBERTa}$, because it does not make sense to perform speaker verification with only semantic embeddings.}
    \vspace{-0.3cm}
    \label{tab:ablation}
\end{table*}

We present the ablation result of different key components in CTAL in Table.~\ref{tab:ablation}. For experimental efficiency, all of the ablation experiments are conducted with $\mathrm{CTAL}_{\mathrm{LARGE}}$.

Overall, the pre-training of CTAL improves the performance across all the three downstream tasks (by comparing settings "w/o Pre-training" and $\mathrm{CTAL}_{\mathrm{LARGE}}$), and we find that $\mathrm{CTAL}_{\mathrm{LARGE}}$ significantly outperforms $\mathrm{CTAL}_{\mathrm{BASE}}$ across all tasks. Besides, with the increment in the size of pre-training data, CTAL achieves better performances on all evaluation metrics except $\rm{Acc_2}$ and F1 in sentiment analysis task (by comparing settings (a) "pre-train with train-clean-360" and $\mathrm{CTAL}_{\mathrm{LARGE}}$). The effectiveness of the asymmetry encoder design for audio-and-language representations is demonstrated by comparing $\mathrm{CTAL}_{\mathrm{LARGE}}$ to $\mathrm{LXMERT}$, where both models are designed to have similar size of parameters.

By comparing (b) "w/o MLM" to "w/o Pre-training" and (c) "w/o MCAM" to "w/o Pre-training", we see the benefits of pre-training on MCAM and MLM respectively. However, by comparing (b) and (c) with $\rm{CTAL_{LARGE}}$, both of them suffer dramatically performance decrease over all downstream tasks. This fact indicates the importance of joint-training with MLM and MCAM task during the pre-training stage. So far, the effectivenesses of pre-training and different tasks are demonstrated. 

Setting (d) "w/o Orthogonal Fusion" removes our proposed cross-modality orthogonal-fusion component and by comparing it with $\mathrm{CTAL}_{\mathrm{LARGE}}$, we observe that the model's performances decrease on all three downstream tasks, which proves its effectiveness. Setting (e) "w/o Audio Outputs" and (f) "w/o Language Outputs" only use the output embeddings from either Audio or Language encoding module for downstream fine-tuning. Through comparing them to (d), we find each of embeddings contributes to the Audio-and-Language tasks and the best performance is achieved through the appropriate fusion of both parts. At last, setting (g) "w/o Cross-modal Pre-training" utilizes unimodal pre-training models, RoBERTa and Mockingjay pre-trained with LibriSpeech dataset, and fuses their output embeddings for the downstream tasks. To be noticed, "w/o Cross-modal Pre-training" is chosen to have the same model size as $\rm{CTAL_{LARGE}}$ for comparison purposes. Additionally, we present the performance of each single modality pre-trained model, $\mathrm{Mockinjay}$ and $\mathrm{RoBERTa}$, to demonstrate the advantages of multimodal pre-training. From the results, we find our approach still holds better performance across all three tasks, which proves the importance of introducing inter-modality learning during pre-training phase.


\subsection{Effect of Pre-training}
We analyze the effect of pre-trained CTAL by visualizing its performance on downstream tasks versus different proportions of training data being used. From the result, CTAL only needs half the amount of training data to achieve better performance than the SOTA baselines. More details are provided in Appendix \ref{sec:effect_of_CTAL}.


\section{Conclusion}
\label{sec:conclusion}
In this work, we proposed CTAL, a novel pre-trainable generic representation for audio-and-language tasks. It is pre-trained with two pre-training tasks on large-scale dataset of audio-and-language pairs. Extensive empirical analysis demonstrates that our pre-trained model can boost various speech understanding performance significantly and achieve new state-of-the-art results. Besides, we show the effectiveness of different model components and the competent generalization capability via detailed ablation studies and analysis.



\bibliographystyle{acl_natbib}
\bibliography{emnlp2021}

\clearpage
\appendix

\section{Appendix}
\label{appendix}
\subsection{Pre-training Details}
\label{sec:pre-train details}
We take the Adam \cite{kingma2014adam} as the optimizer with initial learning rate of 5e-5 and a linear-decayed learning rate schedule with warm up \cite{devlin2018bert}. We pre-train our model using 4 16G-V100 GPUs with a batch size of 16 for 1,000,000 steps, and the whole pre-training process takes roughly 48 hours.

\subsection{Fine-tuning Details}
\label{sec:fine-tune details}
We use a batch size of 4 and fine-tune for 20 epochs over each fold with 1 16G-V100 GPU. We take AdamW \cite{loshchilov2018fixing} as the optimizer in fine-tuning stage, the learning rate is initialized as 1e-5 and we apply a cosine annealing learning rate schedule \cite{loshchilov2016sgdr} to reach the optimum.

\subsection{Baseline for Emotion Classification}
\label{sec:fine-tune baseline}
\citet{xu2019learning} aims to produce more strong multimodal representations by learning the alignment between speech frames and text words using an attention mechanism, we call it LSTM\_alignment in our paper since the original paper did not have a name for their method. MDRE \cite{yoon2018multimodal} uses a dual-RNNs to encode the information from audio and text separately, then combines them by simple representations concatenation to predict emotion classes. MHA \cite{yoon2019speech} proposes a multi-hop attention mechanism to infer the correlation between audio and language modalities based on the output hidden representations of two bi-directional long short-term memory (BiLSTM) encoders, and output the final classification result from the concatenation of audio and language representations.

\subsection{Baseline for Speaker Verification}
\label{sec:fine-tune baseline speaker verification}
GE2E \cite{wan2018generalized} designs a general loss function that emphasizes examples that are difficult to verify at each step of the training process, while RawNet \cite{jung2019rawnet} proposes an end-to-end network that input raw audio waveforms to extract speaker embeddings.

\subsection{Visualizing CTAL Behavior}
\label{sec:effect_of_CTAL}
\begin{figure}
    \small
    \begin{subfigure}{0.495\linewidth}
      \centering
      \includegraphics[width=0.98\linewidth]{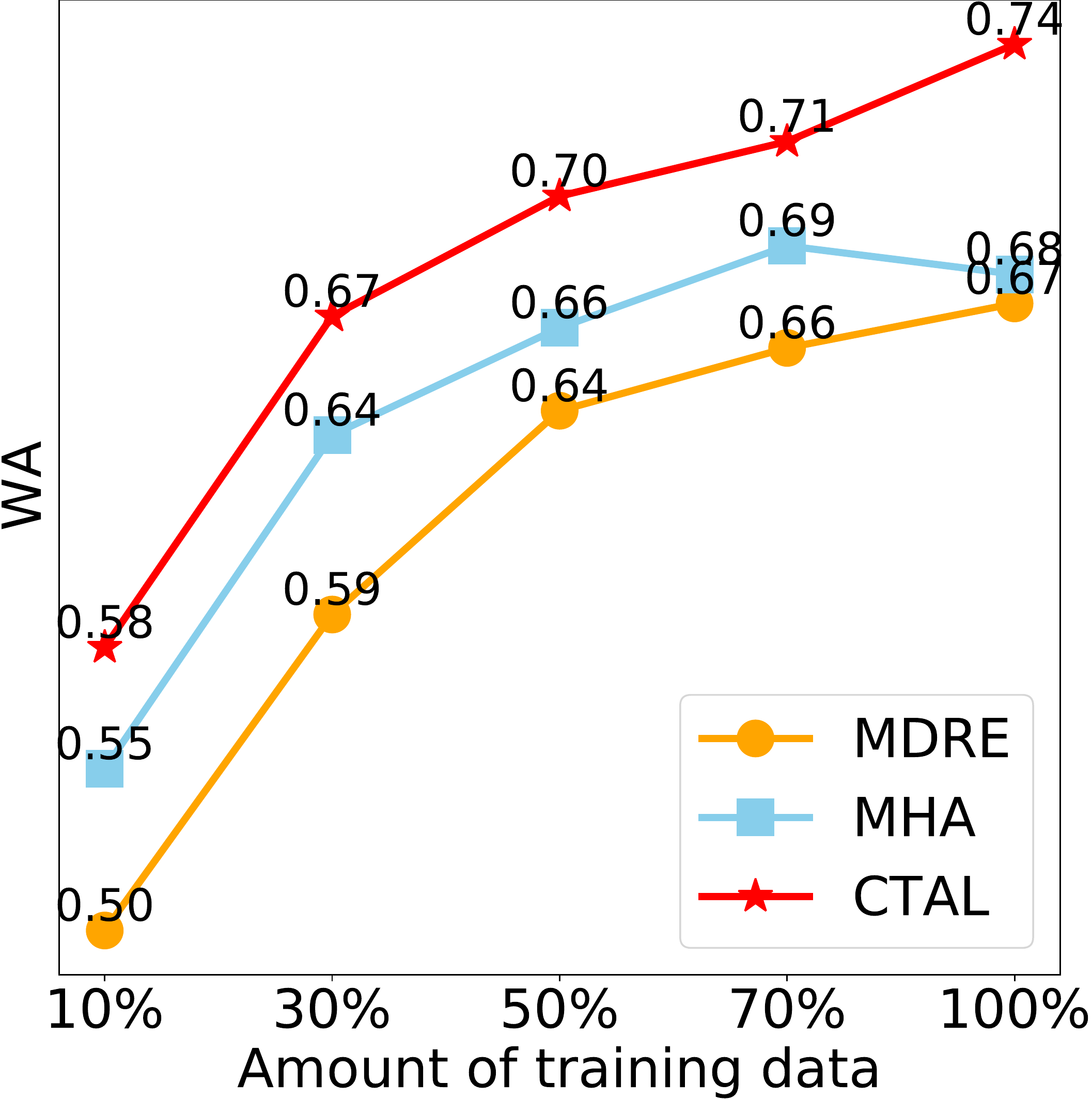}
      \caption{WA vs proportions}
      \label{fig:wa_iemocap}
    \end{subfigure}
    \begin{subfigure}{0.495\linewidth}
      \centering
      \includegraphics[width=0.98\linewidth]{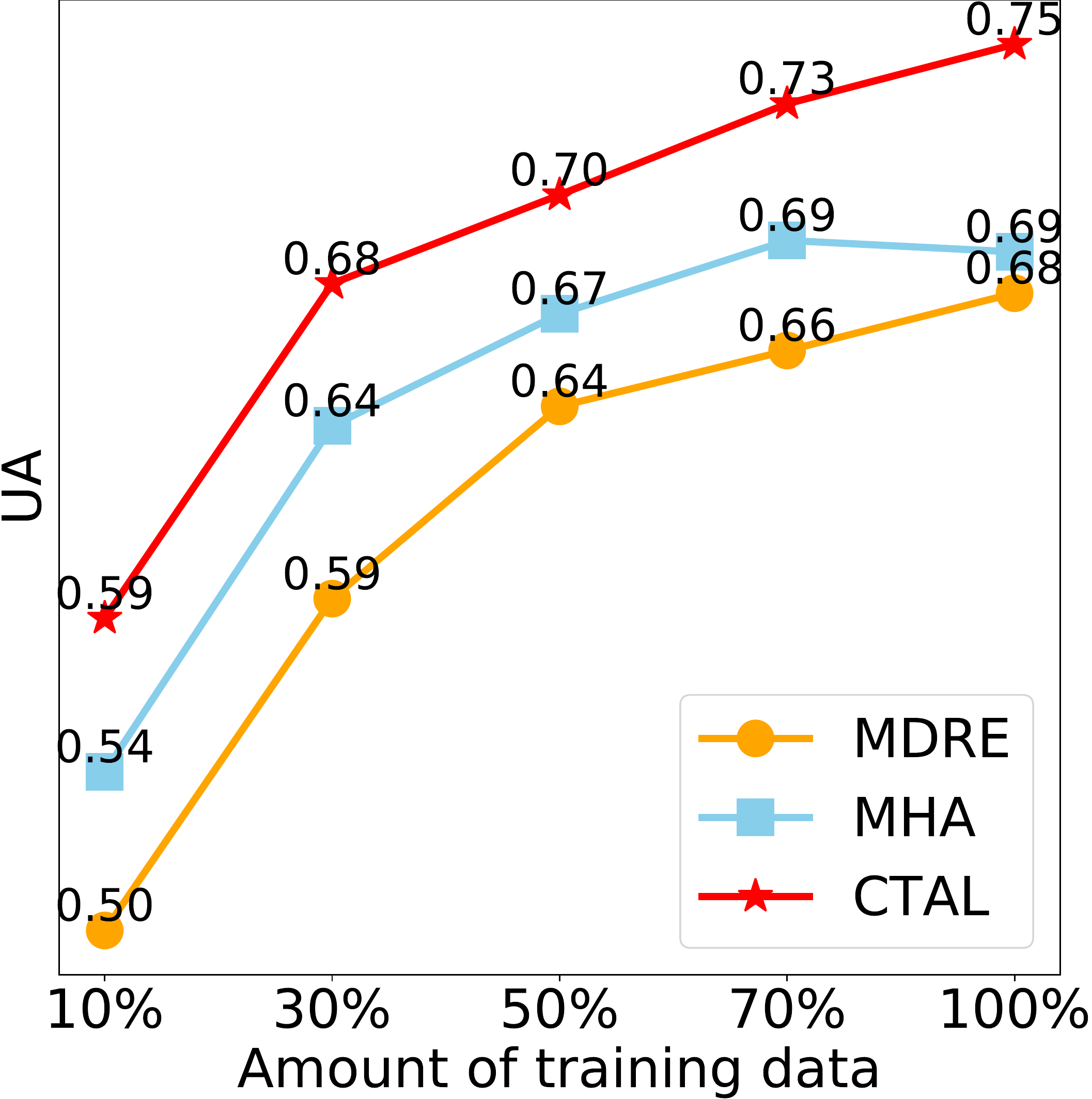}
      \caption{UA vs proportions}
      \label{fig:ua_iemocap}
    \end{subfigure}
    \vspace{-0.25cm}
    \caption{Metrics of different models vs amount of training data on IEMOCAP.}
    \label{fig:wa_ua_iemocap}
\vspace{-0.25cm}
\end{figure}


\begin{figure}
    \small
    \begin{subfigure}{0.495\linewidth}
      \centering
      \includegraphics[width=0.98\linewidth]{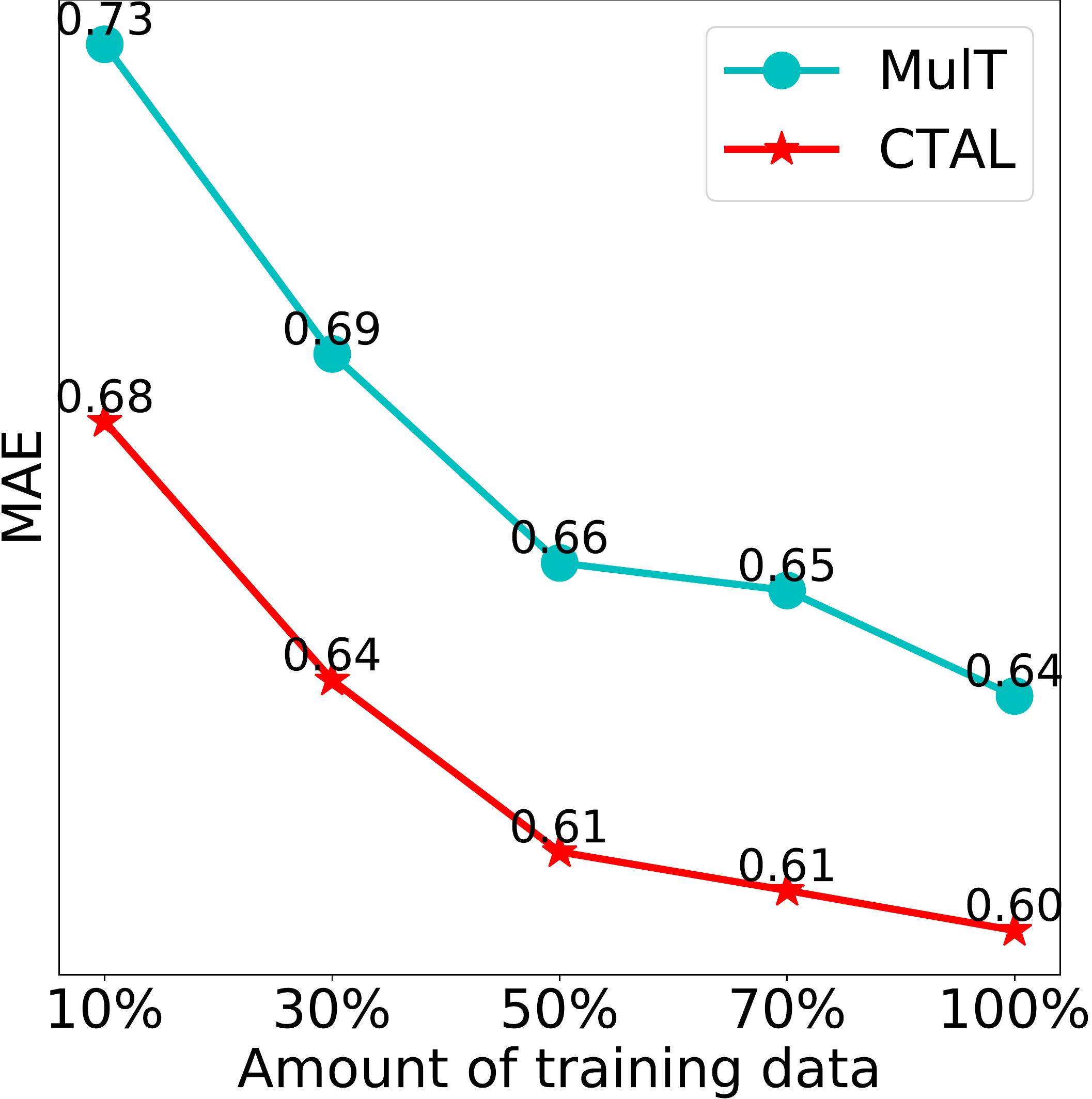}
      \caption{MAE vs proportions}
      \label{fig:mae_mosei}
    \end{subfigure}
    \begin{subfigure}{0.495\linewidth}
      \centering
      \includegraphics[width=0.98\linewidth]{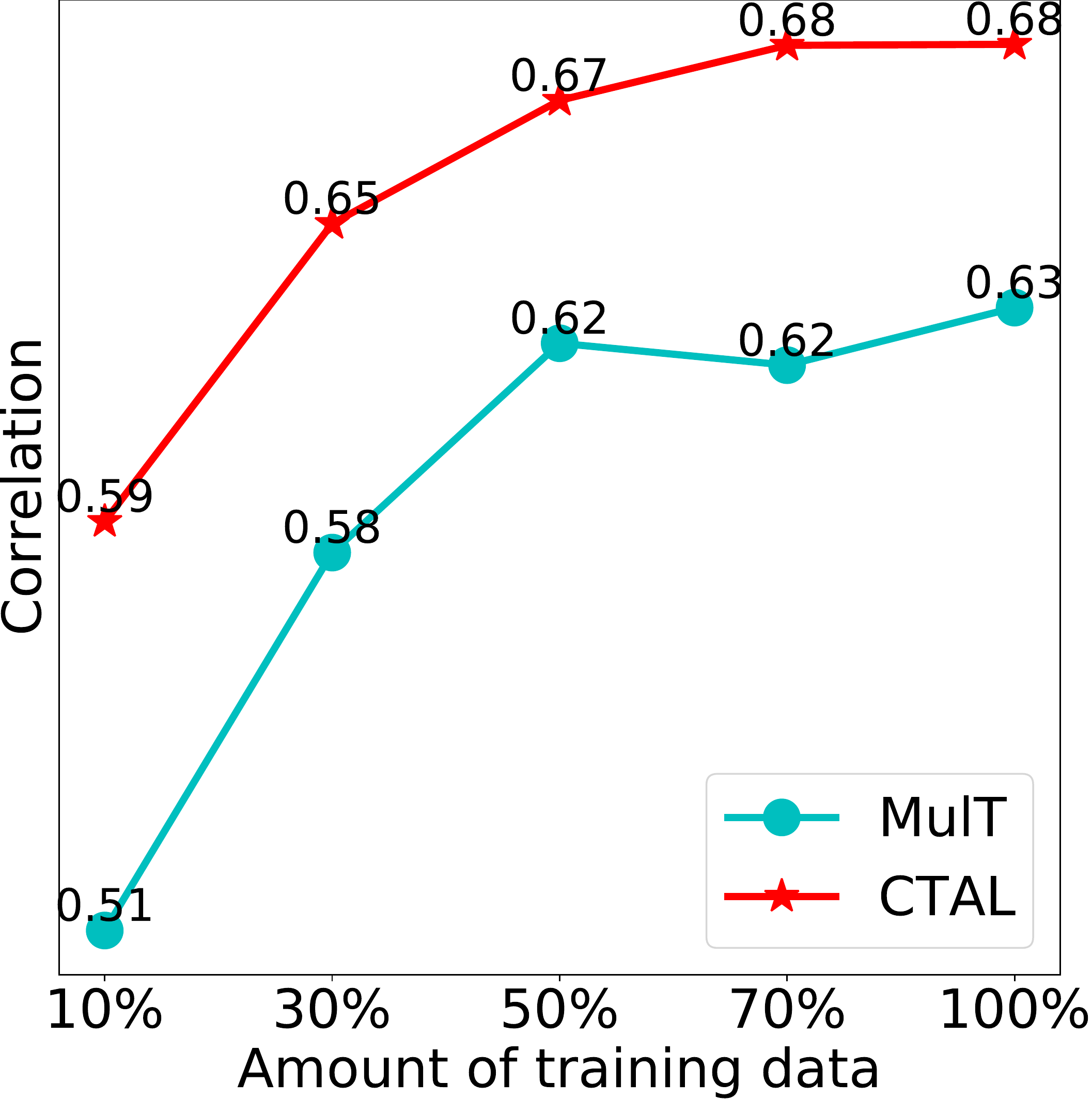}
      \caption{Correlation vs proportions}
      \label{fig:corr_mosei}
    \end{subfigure}
    \vspace{-0.25cm}
    \caption{Metrics of different models vs amount of training data on CMU-MOSEI.}
    \label{fig:mae_corr_mosei}
\vspace{-0.25cm}
\end{figure}

In Figure~\ref{fig:wa_iemocap} and Figure~\ref{fig:ua_iemocap}, we show the performance on IEMOCAP, there are two observations. First of all, on both metrics, CTAL outperforms all baselines across different proportions of training data. Secondly, the figures show that CTAL only needs half the amount of training data to achieve a better performance than baselines. The results on MOSEI are shown in Figure~\ref{fig:mae_mosei} and Figure~\ref{fig:corr_mosei}, and the same conclusion can also be drawn.

\begin{figure}
  \includegraphics[width=0.82\linewidth]{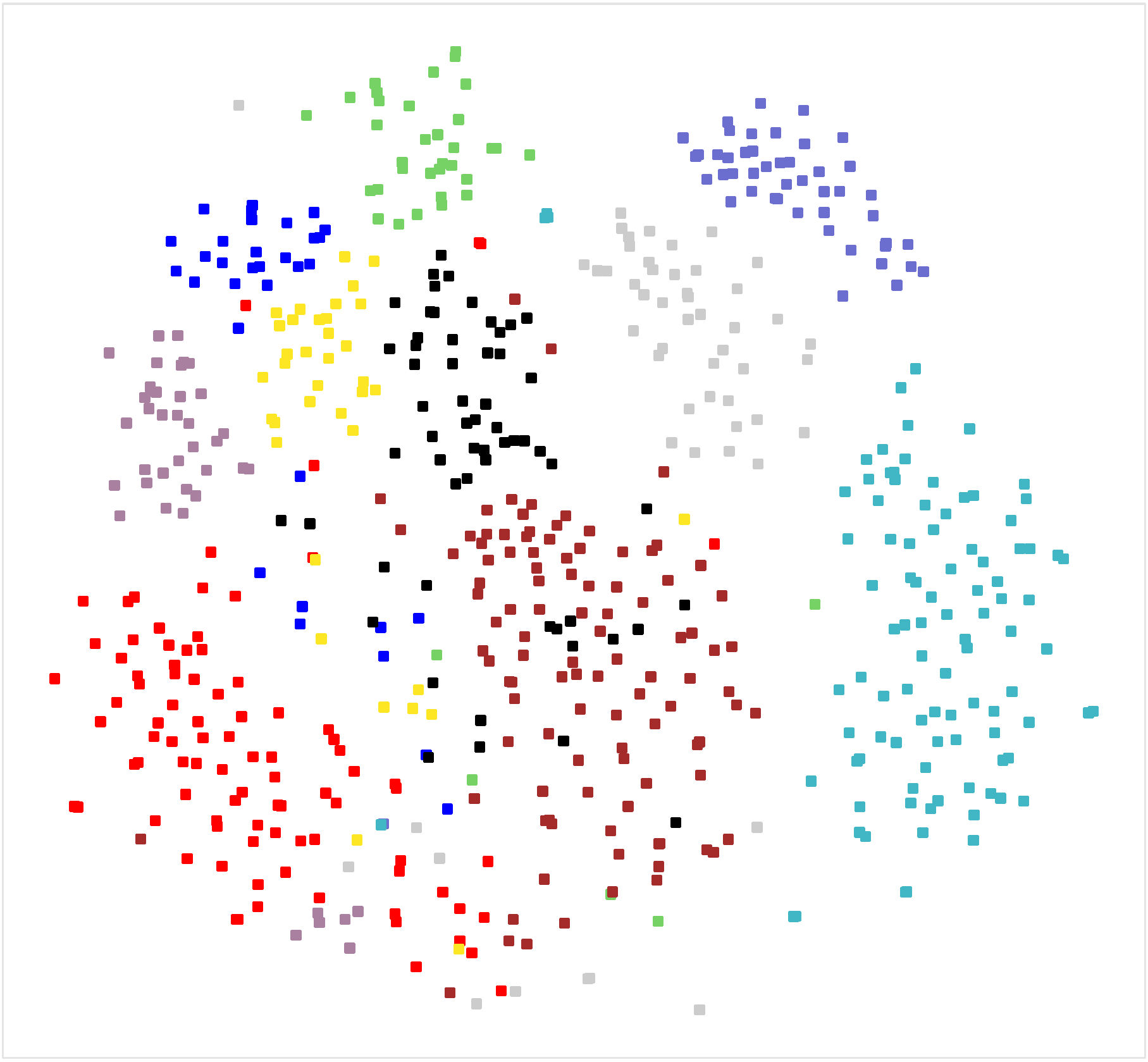}
  \centering
  \vspace{-0.25cm}
  \caption{Visualization of 10 speakers embeddings via t-SNE. Different colors represent different speakers.}
  \label{fig:speaker_embeddings}
  \vspace{-0.5cm}
\end{figure}

In Figure~\ref{fig:speaker_embeddings}, we use t-SNE \cite{van2008visualizing} to visualize the speaker embeddings in test set extracted from pre-trained CTAL without training on downstream tasks. In the figure, each point represents an utterance and different speakers have different colors. We can observe that the model can have some capability to distinguish utterances of different speakers with only pre-training.

\end{document}